\documentclass{article}

\PassOptionsToPackage{numbers}{natbib}
\usepackage[final]{neurips_2019}

\usepackage[utf8]{inputenc} 
\usepackage[T1]{fontenc}    
\usepackage{hyperref}       
\usepackage{url}            
\usepackage{booktabs}       
\usepackage{float}
\usepackage{amsfonts}       
\usepackage{nicefrac}       
\usepackage{microtype}      
\usepackage{graphicx}
\usepackage{natbib}
\usepackage{amsmath,amsthm,amssymb}
\usepackage{xcolor}

\title{March Madness Tournament Predictions Model: A Mathematical Modeling Approach}

\author{
  \textbf{Christian McIver}\thanks{These authors contributed equally to this work.} \\  
  \texttt{cmciver@college.harvard.edu}  
  \and  
  \textbf{Karla Avalos}$^{*}$ \\  
  \texttt{karlaavalos@college.harvard.edu}  
  \and  
  \textbf{Nikhil Nayak}$^{*}$ \\  
  \texttt{nnayak@g.harvard.edu}  
}

\begin{document}

\maketitle






\vspace{-1em}

\begin{abstract}

This paper proposes a model to predict the outcome of the March Madness tournament based on historical NCAA basketball data since 2013. The framework of this project is a simplification of the FiveThrityEight NCAA March Madness prediction model, where the only four predictors of interest are Adjusted Offensive Efficiency (ADJOE), Adjusted Defensive Efficiency (ADJDE), Power Rating, and Two-Point Shooting Percentage Allowed. A logistic regression was utilized with the aforementioned metrics to generate a probability of a particular team winning each game. Then, a tournament simulation is developed and compared to real-world March Madness brackets to determine the accuracy of the model. Accuracies of performance were calculated using a naive approach and a Spearman rank correlation coefficient.

\end{abstract}

\section{Introduction}

	Sport game prediction models have been a topic of interest amongst not just mathematicians, but also the average sports-enjoyer. Knowing the outcome of a match holds many incentives. For some it may be monetary, but for mathematicians and scientists, it may provide more insight into applicative predictive modeling techniques in a broader context. In context to the sport of basketball, at the professional, collegiate, and pre-professional level, modeling the outcome of games based on pre-set conditions has been an inquisitive topic for many. Modeling can help teams adjust and improve on their own strategies to win more games, thus it can be a powerful tool if accurate predictions can be made.
 
	The annual NCAA March Madness tournament is a collegiate basketball tournament that places the top 64 D1 college basketball teams against one another to determine the winner. The tournament consists of (formally) 6 rounds of play, where an initial bracket determined by the relative dominance of a team (a team’s “seed”) is developed. The NCAA tournament bracket is split into four regions of the United States that include the East, West, Midwest, and South. Each region has sixteen teams, which are ranked 1-16 and are the team’s seed as aforementioned. So far, no one in the history of the NCAA March Madness tournament has verifiably predicted the exact outcome of the tournament \cite{NCAA}. Even with prior knowledge of basketball, the estimated odds of predicting a perfect bracket are 1 in 120.2 billion \cite{NCAA}. This project aims to improve those odds by removing natural human subjectivity with decision making by creating a purely objective classification model system to determine the outcome of each game within the tournament.

\section{Model Description}

\begin{figure}[h]
    \centering
    \includegraphics[width=0.3\textwidth]{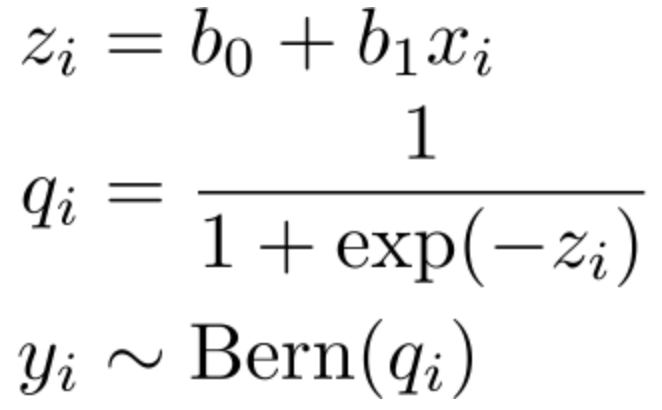}
    \caption{Logistic Regression as a Genralized Linear Model}
    \label{fig:GLM}
\end{figure}

We utilized the logistic regression model to predict the outcome of a match given raw statistics \cite{kaggle} of any two teams participating in the match. Logistic Regression is a type of generalized linear model (see Figure \ref{fig:GLM}).

Generalized linear models consist of three main components:

\begin{enumerate}
    \item \textbf{Linear Predictor:} The linear predictor is the linear combination of the features with specific parameter values given by the model. For logistic regression, the linear predictor is denoted as:
    \[ \eta = \beta_0 + \beta_1 x_1 + \beta_2 x_2 + \dots + \beta_p x_p \]
    where $\eta$ is the linear predictor, $\beta_0, \beta_1, \dots, \beta_p$ are the model parameters, and $x_1, x_2, \dots, x_p$ are the (combined) features of the teams.
    
    \item \textbf{Link Function:} The link function connects the linear predictor to the target variable. In logistic regression, the link function is the logit function, which transforms the linear predictor to the probability scale:
    \[ \text{logit}(p) = \log\left(\frac{p}{1-p}\right) = \eta \]
    where $p$ is the probability of the positive outcome (e.g., winning the match), and $\eta$ is the linear predictor.
    
    \item \textbf{Probability Distribution:} The final component is the probability distribution that generates the observed variable. In logistic regression, since there are only two possible outcomes (e.g., win or loss), it follows a Bernoulli distribution. The probability mass function (PMF) of the Bernoulli distribution for a positive outcome ($y=1$) is given by:
    \[ P(Y = 1 | \mathbf{x}) = p(\mathbf{x}) = \frac{1}{1 + e^{-\eta}} \]
    where $\mathbf{x}$ represents the feature vector of the teams.
\end{enumerate}

The logistic regression model is multivariable since there are more than one explanatory variable, and its description above outlines the fundamental components of logistic regression and their roles in predicting match outcomes based on raw team statistics.

The performance of the model is assessed by testing the predictability of an individual matchup as well as the accuracy of predictability for an entire NCAA bracket. To assess the predictability of an individual matchup we perform a train and test split of the data that includes tournaments for every NCAA tournament from 2013-2023. After model training, the test set is utilized to assess the predictability power on data unseen by the model. To assess the accuracy of predicting an entire tournament, two metrics were utilized. This includes a naive approach that finds the fraction of tournaments predicted accurately over the total number of games in the tournament. The second metric utilizes the Spearman rank correlation to compare the final round or “rank” of each team in the simulation and in the data. The Spearman rank correlation, $\rho$, measures the strength and direction of association between two ranked variables and is given below. In this case, the ranked variables are the final round in the tournament for each team in the simulation and in the real data. 

\[ \rho = 1 - \frac{6 \sum d_i^2}{n (n^2-1)} \]

where $d_i$ is the difference in ranks across the two variables, and $n$ is the number of observations. 

\section{Analysis}

In developing our multivariable logistic regression model to predict match outcomes, we encountered several modeling considerations that were crucial for achieving a robust predictive model. We address each of these considerations below:

\begin{enumerate}
    \item \textbf{Handling Missing Data:} We encountered missing data in the Kaggle dataset used for training our model. To address this, we employed mean value imputation. Specifically, we replaced missing values for a particular team with the mean value of that feature from other games in the same year. Subsequently, any rows with remaining missing values were dropped from the dataset.
    
    \item \textbf{Feature Engineering:} Rather than using both teams' raw feature values directly, we opted to utilize the difference between the two teams' feature values as features. This approach effectively reduces the number of features while still capturing relevant information for prediction. This has the advantage of explicitly modeling the probability $p$ of team A winning when input to the logistic function is $x$ and probability $1-p$ of team B winning when the input is $-x$.
    
    Let's denote the logistic function as $f(x) = \frac{1}{1 + e^{-x}}$.
    
    When $x$ is input to the logistic function, it gives probability $p$:
    \[ f(x) = \frac{1}{1 + e^{-x}} = p \]
    
    And when $-x$ is given, it gives probability $1-p$:
    \[ f(-x) = \frac{1}{1 + e^{x}} = 1-p \]
    
    \item \textbf{Standardization:} Standardization of features is essential to ensure that all features have the same scale (mean 0 and standard deviation 1). The equation for data standardization in logistic regression is given by:
    \[
    x_{\text{standardized}} = \frac{x - \mu}{\sigma}
    \]
    
    where:
    \begin{itemize}
        \item $x_{\text{standardized}}$ is the standardized version of the original feature $x$,
        \item $x$ is the original feature,
        \item $\mu$ is the mean of the feature $x$ in the training data,
        \item $\sigma$ is the standard deviation of the feature $x$ in the training data.
    \end{itemize}

    This preprocessing step facilitates regularization and prevents any particular feature from being unfairly penalized during model training.
    
    \item \textbf{Regularization:} To prevent overfitting and ensure that coefficients of features are kept small, we employed L2 regularization. This technique adds an L2 norm term penalty to the loss function, effectively penalizing large coefficients and encouraging sparsity in the model. The updated cost function is given by:
    \[
    J(\theta) = \frac{1}{m} \sum_{i=1}^{m} \left[-y^{(i)} \log(h_\theta(x^{(i)})) - (1 - y^{(i)}) \log(1 - h_\theta(x^{(i)}))\right] + \lambda \sum_{j=1}^{n} \theta_j^2
    \]

    where \( J(\theta) \) is the regularized cost function, \( m \) is the number of training examples, \( n \) is the number of features, \( y^{(i)} \) is the target variable for the \( i \)th training example, \( h_\theta(x^{(i)}) \) is the hypothesis function, \( x^{(i)} \) is the feature vector for the \( i \)th training example, \( \theta \) is the parameter vector, \( \lambda \) is the regularization parameter (also known as the regularization strength). The regularization parameter controls the strength of regularization, balancing between fitting the training data well and generalizing to unseen data.

\end{enumerate}

These modeling considerations were crucial in developing a predictive model that effectively captures the underlying patterns in the data while mitigating issues such as missing data and overfitting.

After utilizing the selected logistic regression model, Monte Carlo simulations were run to simulate an entire tournament. Data for the initial matchups for the NCAA 2023 and 2022 tournament were utilized to initiate the simulation. For a particular matchup, the differences between Adjusted Offensive Efficiency, ADJOE, Adjusted Defensive Efficiency, ADJDE, Power Rating, BARTHAG, and Two-Point Shooting Percentage Allowed, 2P$_D$, were utilized to find the probability of each team winning. A randomly generated value determined the winner for each matchup for each round. This process was repeated until a winner was determined for each tournament. Below is an example of one iteration of a simulation on the 2023 tournament for each conference bracket. The predicted overall champion by our model for the 2023 tournament is University of Houston.

\begin{figure}[h]
    \centering
    \includegraphics[width=0.7\textwidth]{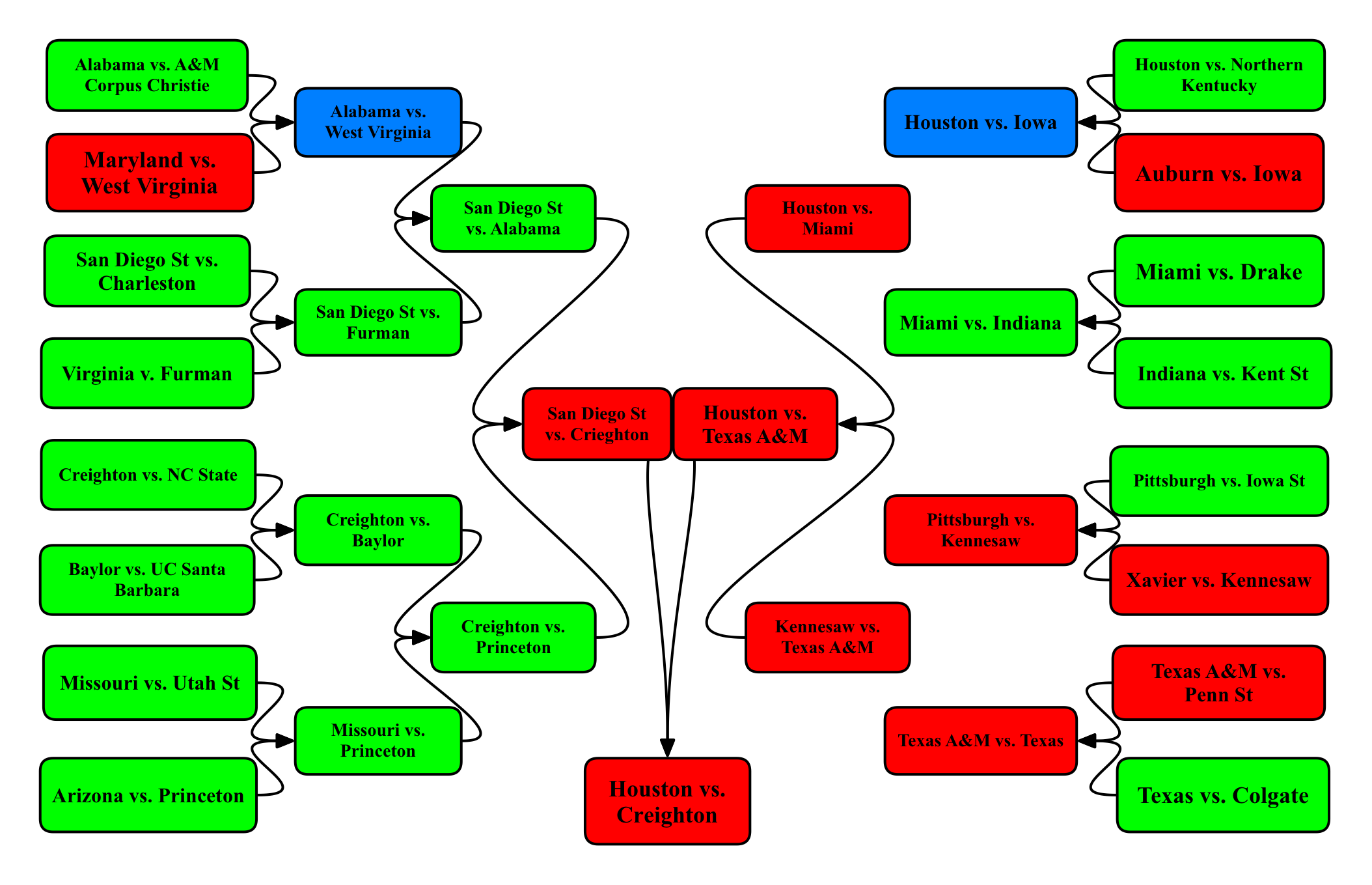}
    \caption{Simulation results for the NCAA 2023 tournament between the South and Midwest conferences, on the left and right side of the bracket, respectively. The green matchups represent matchups correctly predicted by the logistic regression model. The red matchups represent matchups that were predicted incorrectly. The blue matchups represent matchups that were composed of the wrong teams but had the correct winner.}
    \label{fig:SouthvMid}
\end{figure}

\begin{figure}[h]
    \centering
    \includegraphics[width=0.7\textwidth]{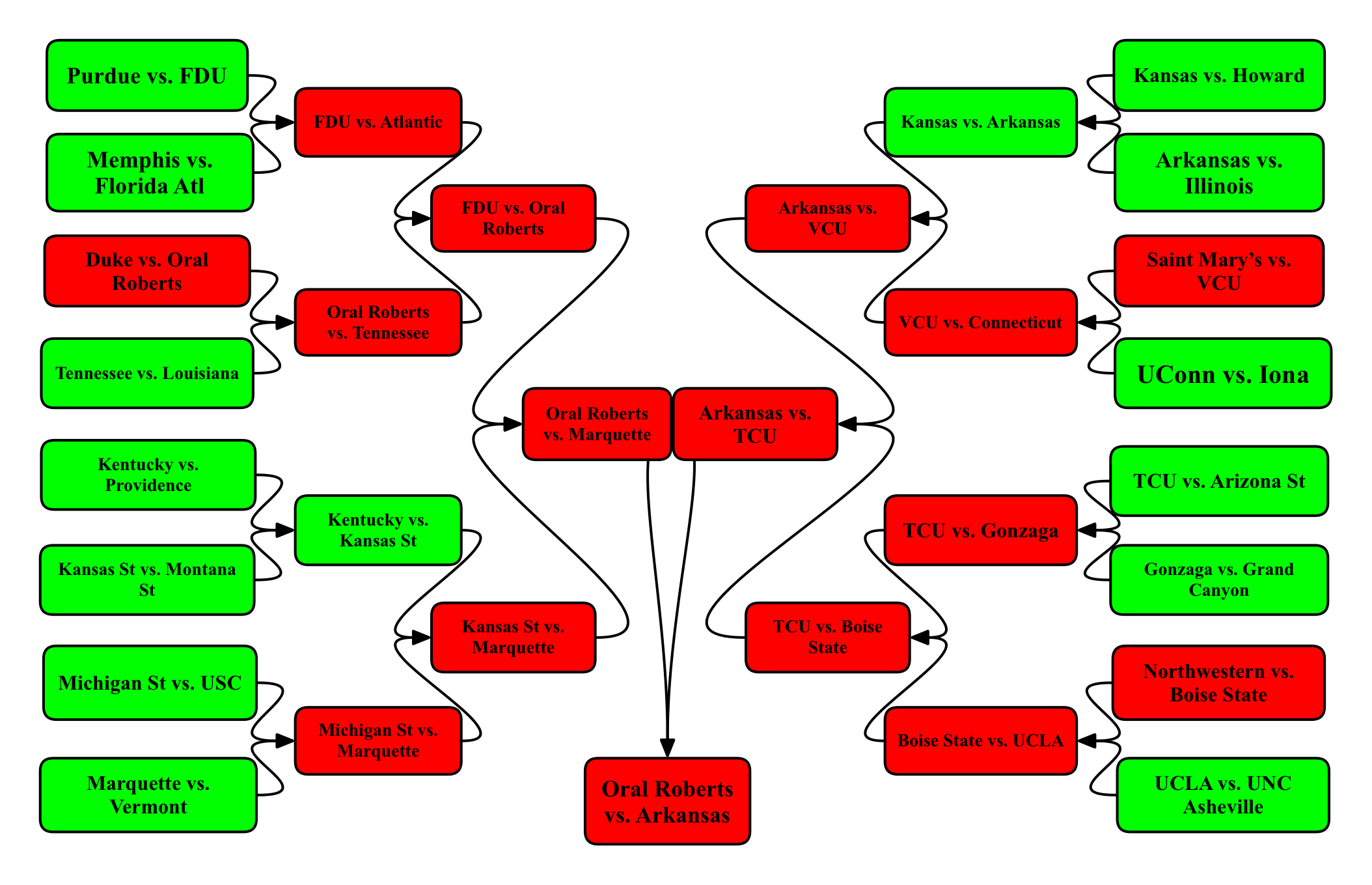}
    \caption{Simulation results for the NCAA 2023 tournament between the East and West conferences, on the left and right side of the bracket, respectively.}
    \label{fig:EastvWest}
\end{figure}

A naive approach to assessing the performance of the model over an entire bracket is finding the fraction of matchups predicted correctly. Each regional tournament consists of 31 matchups prior to the championship matchup. Matchups that correspond to games that were composed of the wrong teams but had the correct winner, labeled in blue above, are only weighted as “half correct” when evaluating the fraction of matchups predicted correctly. Table \ref{tab:simulation_accuracy} summarizes the metric for the simulation of the 2023 and 2022 tournaments over 100 iterations. 

To assess the performance of the model over many iterations, a Spearman correlation coefficient was also calculated for the tournament. A Spearman correlation coefficient was utilized to assess the strength and direction of association between the predicted final round and the actual final round for each team. This approach effectively utilizes the final round of each team as the rank of the team. In a scenario where our model perfectly predicts every single match, we can expect a Spearman correlation coefficient of one since there would be no difference in the final rounds for each team in the simulation and real data. 

Both brackets pictured from figure \ref{fig:SouthvMid} and \ref{fig:EastvWest} were simulated for 100 iterations and the corresponding Spearman correlation coefficients are summarized below. The results suggest that our model predicted the South vs. Midwest bracket better than the East vs. West bracket. The Spearman correlation coefficient values are consistent with the visual inspection of the brackets, with the bracket in figure \ref{fig:SouthvMid} having a greater amount of matchups predicted accurately. Both Spearman correlation coefficients indicate a moderate to strong positive association between the predicted and actual final rounds for each team.

\begin{table}[h]
    \centering

    \caption{Resulting naive accuracies and Spearman correlation coefficients for the simulated tournaments. }
    \label{tab:simulation_accuracy}
    \begin{tabular}{lcc}
        \toprule
        \textbf{Bracket Simulation} & \textbf{Games Predicted} & \textbf{Spearman Correlation} \\
         & \textbf{Correctly (\%)} & \textbf{Coefficient ($\rho$)} \\
        \midrule
        2023 South vs. Midwest & 65.63 & 0.7467 \\
        2023 East vs. West & 43.75 & 0.4875 \\
        2022 South vs. West & 56.25 & 0.6311 \\ 
        2022 East vs. Midwest & 50 & 0.3651\\
        \bottomrule
    \end{tabular}
\end{table}
  
\section{Discussion}

\begin{figure}[h]
    \centering
    \includegraphics[width=0.7\textwidth]{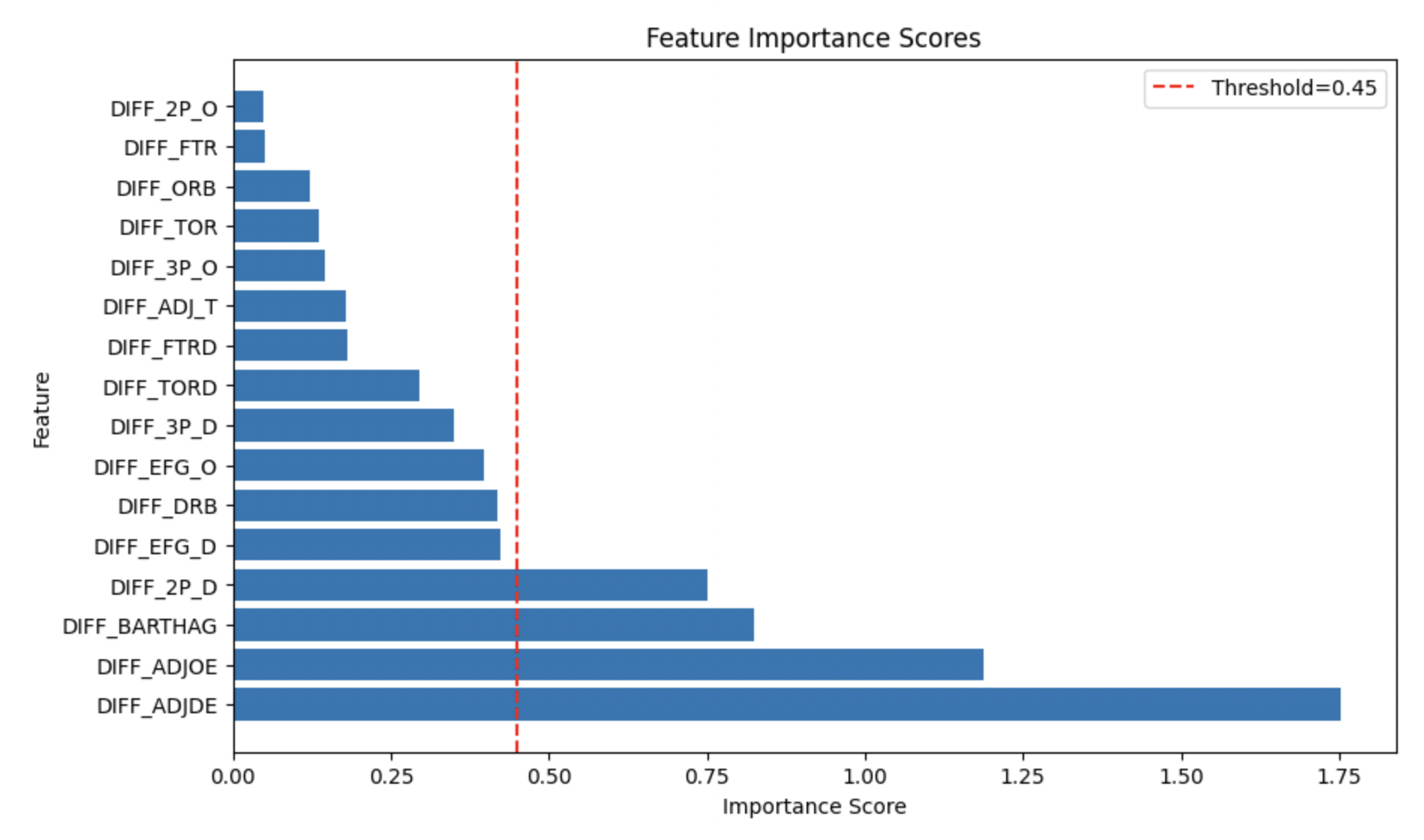}
    \caption{Feature Importance Plot}
    \label{fig:FeatureImportance}
\end{figure}

The multivariable logistic regression model utilized a total of 16 explanatory variables. For feature selection, we examined the most important features of the model based on the magnitude of their coefficients. The feature importance plot is shown in Figure \ref{fig:FeatureImportance}. Features with coefficients below the specified threshold (set to 0.45 after analyzing the feature importance plot) were discarded. This approach was adopted to mitigate overfitting and address multicollinearity, thereby enhancing model interpretability. The important features identified through this process included the adjusted defensive efficiency (i.e., the number of points allowed per 100 possessions), adjusted offensive efficiency (i.e., the number of points scored per 100 possessions), power rating, and 2-point shooting percentage allowed.

To assess the impact of feature selection on model performance, we compared the training and testing accuracies of the original model with those of the model after feature selection. The results are presented in Table~\ref{tab:accuracy_comparison}.

\begin{table}[h]
    \centering
    \caption{Training and Testing Accuracy Comparison}
    \label{tab:accuracy_comparison}
    \begin{tabular}{lcc}
        \toprule
        \textbf{Model} & \textbf{Training Accuracy (\%)} & \textbf{Testing Accuracy (\%)} \\
        \midrule
        Original Model & 72.76 & 75.39 \\
        Model after Feature Selection & 71.35 & 74.60 \\
        \bottomrule
    \end{tabular}
\end{table}

From the table, we observe that the model after feature selection demonstrates slightly lower training and testing accuracies compared to the original model. However, these differences are marginal, indicating that the selected features effectively capture the predictive power of the model while reducing complexity. Also, in both the original and retrained models, the test accuracy is higher than the train accuracy likely due to the use of regularization, which helps prevent overfitting to the training data.

\begin{figure}[h]
    \centering
    \includegraphics[width=0.7\textwidth]{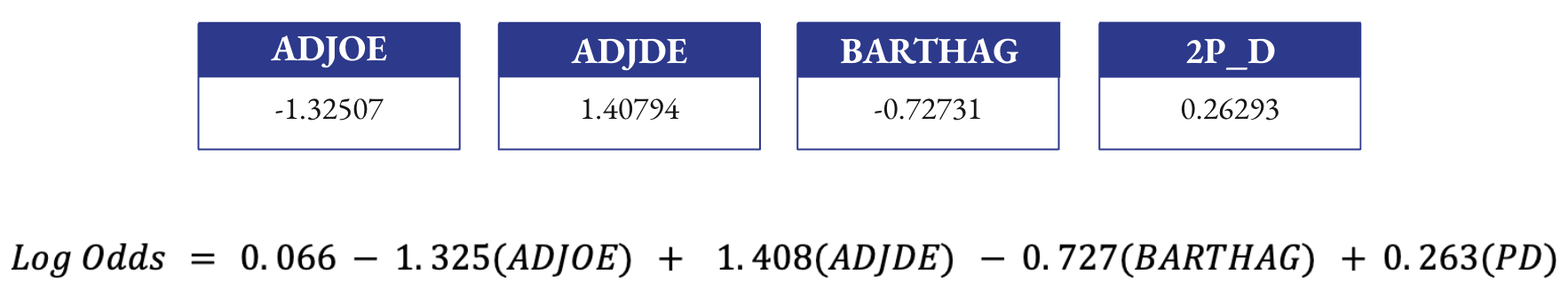}
    \caption{Selected Model Equation}
    \label{fig:SelectedModel}
\end{figure}

The final model equation along with selected features and their coefficients are shown in Figure \ref{fig:SelectedModel}. The behavior of the model with respect to the selected features aligns with our expectations. Features such as adjusted defensive and offensive efficiencies, power rating, and 2-point shooting percentage allowed are intuitive indicators of team performance and are likely to influence match outcomes. By focusing on these key features, our model enhances interpretability and provides valuable insights into the factors driving match predictions. One noteworthy aspect learned from the model is the importance of defensive and offensive efficiencies in determining match outcomes. These metrics encapsulate a team's ability to both score points efficiently and prevent their opponents from doing so, highlighting their significance in basketball analytics.

Despite the strengths of our analysis, there are some limitations to consider. The model's performance may be influenced by factors not captured in the dataset, such as team strategies, player injuries, and external factors like venue and crowd dynamics. Additionally, while feature selection aids in reducing model complexity, it may overlook potentially relevant variables, necessitating further exploration in future studies.

The implemented logistic regression model provides valuable insights in the development of robust predictive models for tournament prediction and simulation. It has direct interpretability for each predictor and its corresponding effects to the logs odds of winning for a particular team. Although a test accuracy score of 74.6 percent for individual matchups are promising, the simulations of entire tournament brackets show predictive power but there are inconsistencies that merit further investigation. For example, in table \ref{tab:simulation_accuracy} it can be seen that within the same tournament year there are large discrepancies in the fraction of games predicted correctly and the Spearman correlation coefficient within the regional brackets. These discrepancies may be an indication of bias within our model that was not correctly accounted for either during the train and test split or via L2 regularization. 

Additionally, it is worth noting that the raw statistics, such as adjusted offensive efficiency, utilized in training the model and predicting tournament matchups are assumed to be constant throughout an individual year. This assumption imposes limitations on the efficacy of the model as a team’s statistics can rapidly change during a season or tournament as players get injured, game strategies evolve, and opponents change. 

Nonetheless, for bracket simulations such as the 2023 South vs. Midwest or the 2022 South vs. West, we observe that the fraction of games predicted is better than 50 percent with a moderate to strong relationship between the predicted and actual final rounds of a team. Utilizing raw statistical data to robustly predict tournaments appears feasible. Based on the two simulated tournaments, it appears that some of the bracket simulations are better than a random coin flip and are comparable to more complex predictive models such as FiveThirtyEight’s predictive model utilizing computer ratings and human rankings \cite{538}. 

\section{Summary}

The initial aim of the project to generate a purely objective classification model to determine the outcome of March Madness tournaments was met. Via feature selection, four primary basketball attributes were utilized (ADJOE, ADJDE, Power Rating, and Two-Point Shooting Percentage Allowed) as inputs to a logistic regression model with outputs being log-odds probabilities of a particular team winning in a 1-on-1 match. Such features were able to achieve a 74.6\%  testing accuracy, with insights to the particular significance of ADJOE and ADJDE on match outcomes based on their final model coefficients. The log-odds probabilities collected were then employed in Monte Carlo simulations with 2022 and 2023 initial March Madness brackets. Though results varied widely, with the naive fraction of games predicted correctly ranging from 0.56 to 0.65 and Spearman rank coefficients ranging from 0.365 to 0.746, we are still confident that raw statistics have merit in generally predicting match outcomes. Future steps for improvement include adjusting limitations in the model, such as removing the assumption that predictors remain constant throughout the season and perhaps "fluctuate" as a function of other inputs, increasing complexity but not losing the generality and interpretability of the model's outcomes. 

\section{Attribution of Effort}

Christian, Karla, and Nikhil came up with the problem statement. Christian, Karla, and Nikhil came up with the key ideas of the main methodology. Nikhil implemented the logistic regression model and the feature importance computation. Christian and Karla implemented the monte carlo simulation and tournament bracket simulation. Christian, Karla, and Nikhil led the development of theoretical results. Christian and Karla also identified appropriate datasets and Christian, Karla, and Nikhil carried out the necessary preprocessing of these datasets. Christian, Karla, and Nikhil designed and implemented the experimental evaluation. Christian, Karla, and Nikhil did the writing of the final project report. Christian, Karla, and Nikhil wrote and documented the code in the ipython notebook file. All the authors also helped each other with brainstorming as well as double-checking their work.

\medskip

\bibliographystyle{unsrt}
\nocite{*}
\bibliography{final.bib}

\end{document}